\documentclass[fleqn,twoside]{article}
\usepackage[headings]{espcrc2}

\usepackage{graphicx}
\usepackage[figuresright]{rotating}

\newcommand{\etal}{\textit{et al.}}
\newcommand{\plb}{Phys.~Lett.~B}
\newcommand{\prl}{Phys.~Rev.~Lett.}
\newcommand{\prd}{Phys.~Rev.~D}
\newcommand{\zpc}{Z.~Phys.~C}
\newcommand{\epjc}{Eur.~Phys.~J.~C}

\newcommand{\pb}{pb$^{-1}\ $}
\newcommand{\fb}{fb$^{-1}\ $}
\newcommand{\qsq}{\ensuremath{q^2}}
\newcommand{\gevcsq}{\ensuremath{\textrm{GeV}/c^2}}

\newcommand{\dztoxen}{\ensuremath{D^0 \to X e^+ \nu_e}}
\newcommand{\dtoxen}{\ensuremath{D^+ \to X e^+ \nu_e}}
\newcommand{\dztokln}{\ensuremath{D^0 \to K^- \ell^+ \nu_\ell}}
\newcommand{\dztokmn}{\ensuremath{D^0 \to K^- \mu^+ \nu_\mu}}
\newcommand{\dztoken}{\ensuremath{D^0 \to K^- e^+ \nu_e}}
\newcommand{\dztopiln}{\ensuremath{D^0 \to \pi^- \ell^+ \nu_\ell}}
\newcommand{\dztopimn}{\ensuremath{D^0 \to \pi^- \mu^+ \nu_\mu}}
\newcommand{\dztopien}{\ensuremath{D^0 \to \pi^- e^+ \nu_e}}
\newcommand{\dtokzen}{\ensuremath{D^+ \to K^0 e^+ \nu_e}}
\newcommand{\dtopizen}{\ensuremath{D^+ \to \pi^0 e^+ \nu_e}}
\newcommand{\dtokrln}{\ensuremath{D^+ \to \overline{K}^{*0} \ell^+ \nu_\ell}}

\newcommand{\dtokpiln}{\ensuremath{D^+ \to K^-\pi^+ \ell^+ \nu_\ell}}

\newcommand{\thv}{\ensuremath{\theta_{V}}}
\newcommand{\thl}{\ensuremath{\theta_\ell}}
\newcommand{\costhv}{\ensuremath{\cos\thv}}

\newcommand{\sinthv}{\ensuremath{\sin\thv}}
\newcommand{\costhl}{\ensuremath{\cos\thl}}

\newcommand{\sinthl}{\ensuremath{\sin\thl}}

\newcommand{\Hpls}{\ensuremath{H_+(\qsq)}}
\newcommand{\Hmin}{\ensuremath{H_-(\qsq)}}
\newcommand{\Hzer}{\ensuremath{H_0(\qsq)}}
\newcommand{\hzer}{\ensuremath{h_0(\qsq)}}
\newcommand{\Hspls}{\ensuremath{H^2_+(\qsq)}}
\newcommand{\Hsmin}{\ensuremath{H^2_-(\qsq)}}
\newcommand{\Hszer}{\ensuremath{H^2_0(\qsq)}}
\newcommand{\Hint}{\ensuremath{\hzer\,\Hzer}}

\title{A review of charm semileptonic decays}

\author{Doris Yangsoo Kim\address{University of Illinois at Urbana-Champaign,
        1110 W Green St., Urbana, IL. 61801, US}
}
       
\runtitle{A review of charm semileptonic decays}
\runauthor{D. Kim}

\begin{document}

\begin{abstract}
Several high energy experiments have been actively pursuing the analyzes of
semileptonic decays of charm mesons, resulting in numerous new results and
publications. In this report, we summarize the recent efforts on the topics
of pseudoscalar and vector charm semileptonic decays, especially in the area
of branching fractions and form factor measurements.
\end{abstract}

\maketitle

\section{INTRODUCTION}

The semileptonic decays of charm particles continue to be an important 
area of research. The branching fractions provide
measurements of CKM matrix elements in the charm sector, $|V_{cq}|^2$.
All hadronic
complications are contained in the form factors, which can be calculated
via non-perturbative lattice QCD, Heavy Quark Effective Theories, Quark Models,
and other methods. By comparing experimental observations with the
lattice QCD calculations, we obtain a high quality lattice calibration, which
is crucial in reducing systematic uncertainties in the unitarity triangle. The
same technique validated by charm decays can be applied to beauty decays,
subsequently improving the CKM matrix elements in the beauty sector.

Several experiments study charm particle decays, including
charm and $B$- factories and fixed target experiments.
The CLEO-c experiment started taking data recently and has collected
281 \pb at $\Psi(3770)$. Since this energy is
the threshold for the $D^+D^-$ and $D^0\overline{D^0}$ production,
the number of decay particles in the detector is relatively small,
the event reconstruction environment is quite clean, particle
identification is excellent, and backgrounds are negligible.
Last year, they reported
branching fraction measurements of semileptonic charm decays using
the first 56 \pb sample~\cite{cleoexcl}. Their results are already comparable or
better than the PDG 2004 world averages~\cite{pdg2004}.

\section{INCLUSIVE SEMILEPTONIC BRANCHING FRACTIONS}

CLEO-c (06)~\cite{cleoincl} reported both new measurements on the inclusive
branching fractions of \dztoxen{} and \dtoxen{} decays and the corresponding
electron momentum spectra based on the 281 \pb sample collected
at the $\Psi(3770)$.
At least 200~MeV/$c$ is required for the electron momentum.
The $D^0$ and $D^+$ spectra look similar to each other, as expected by
theories.  The spectra are extrapolated for the lower momentum region
to obtain the branching fraction numbers.  The results are:
   (6.46 $\pm$ 0.17 $\pm$ 0.13) \% and 
   (16.13 $\pm$ 0.20 $\pm$ 0.33) \%
for the branching fractions of the \dztoxen{}  and \dtoxen{} decays,
respectively.  The sums of the known exclusive semielectronic decay modes are
   (6.1 $\pm$ 0.2 $\pm$ 0.2) \% for $D^0$ and
   (15.1 $\pm$ 0.5 $\pm$ 0.5) \% for $D^+$.
Hence, the sums of the exclusive branching fractions come close
to saturating the inclusive branching fractions, leaving little 
room for any undiscovered semileptonic decays.

\section{A FADING ENIGMA ON THE RATIO OF VECTOR TO PSEUDOSCALAR DECAY RATES}

There has been a disagreement on the decay width ratio of 
the vector semileptonic decays relative to the pseudoscalar semileptonic decays:
$\Gamma(D \to K^* \ell \nu_\ell)/ \Gamma(D \to K \ell \nu_\ell)$.
The earlier theoretical predictions for this ratio from late 80's and early 90's
were found to be between 0.5 to 1.1, which are 50 to 100 \% larger than the
experimental measurements at that time. This difference between the theory
and the data was attributed to difficulties in computing the axial-vector form
factor ($A_1$). Since mid 90's, the updated branching ratio predictions
are between 0.5 to 0.7, being closer to the experimental results.

Recently, several experiments revisited this vector-pseudoscalar width ratio
problem. FOCUS (04)~\cite{will} and CLEO (05)~\cite{cleoexcl} reported
this ratio as $0.594 \pm 0.05$ and $0.63 \pm 0.04$, respectively.
This year BES~\cite{bes1,bes2} reported two measurements: 
$0.57 \pm 0.17$ for the $D^+$ channel and $0.74 \pm 0.40$ (preliminary)
for the $D^0$ channel. These new sets of measurements are
consistent with one another and with the updated predictions.
The old mismatch in the width ratio appears to be fading away.   
   
\section{FORM FACTORS OF PSEUDOSCALAR DECAYS}

The differential decay rate for a $D$ meson decaying into a pseudoscalar meson,
a lepton and a neutrino is given by Eq.~\ref{rate}.
\begin{equation}
  \frac{d\Gamma(D \to P \ell \nu_\ell)}{d\qsq} =
      \frac{G_F^2 |V_{cq}|^2 P_P^3}{24 \pi^3}
      {|f_+(\qsq)|^2 + O(m_\ell^2)}, \label{rate}
\end{equation}
where \qsq{} is the invariant mass of the lepton-neutrino pair.
The pseudoscalar semileptonic process should provide both 
a clean measurement of CKM angles and a powerful test of lattice QCD.
Unfortunately, the QCD test is compromised somewhat since
the decay rate vanishes at the highest \qsq{} where the
sensitivity to the form of $f_+(\qsq)$ is greatest. The highest \qsq{}
area is also the zero recoil limit region for the quark inside the meson
where theory calculations can be obtained relatively easy. 

\subsection{Parameterization of form factor measurements}

To handle the measured \qsq{} distributions properly, several fit
parameterization have been suggested. Based on the dispersion relation,
the form factor $f_+(\qsq)$ can be described by a pole term and an integral
such as,
\begin{equation}
  f_+(\qsq) = \frac{\mathcal{R}}{m_{D^*}^2-\qsq} +
   \frac{1}{\pi}\int_{(m_D+m_P)^2}^{\infty} 
   \frac{\mathrm{Im} f_+(s)}{s-\qsq-i\varepsilon} ds,
\end{equation} 
where $m_{D^*}$, $m_D$ and $m_P$ are the masses of the excited spin-1 $D$ meson,
the decaying $D$ meson, and the pseudoscalar meson, respectively. In the past,
only a ``single" pole term was used to fit the data distribution. But the 
world average of the fitted pole mass for the \dztokln{} decay 
turned out to be 5.1 $\sigma$ lower than the expected value, $m_{D_S^*}$,
underscoring the importance of the integral term.

Becirevic and Kaidalov~\cite{bec} represented the integral by an additional,
effective pole which leads to a parameterization of 
$f_+(\qsq) = f_+(0)/\{(1-\qsq/m_{D^*}^2)(1-\alpha\,\qsq/m_{D^*}^2)\}$. 
This scheme is called the modified pole form.
Recently, Hill~\cite{hill} proposed a less model dependent way of dealing with
the $f_+$ analytic singularities. He makes a complex mapping from \qsq{} to $z$,
which pushes the cut singularities far from the physical \qsq{} region.
In the $z$ space, the $f_+(z)$ distribution looks linear and can be represented
by a rapidly converging Taylor series. He applied the method to the recent
beauty and charm data sets. The results are given in Reference~\cite{hill}. 

\subsection{Experimental results from BABAR, BELLE, CLEO, and FOCUS}

The current generation of experiments have produced new data sets,
obtaining parametric fits to various \qsq{} forms. In addition,
some experiments presented the \qsq{} distributions model-independently, i.e., non-parametrically,
making the visible comparison to the most recent lattice QCD results (05)
possible~\cite{lqcd}.  The $\alpha$ measurements from
present experiments are summarized in Table~\ref{alpha}.

FOCUS (05) measured the form factors of the \dztokmn{} and \dztopimn{}
decays and presented the model-independent $f+(\qsq)$ distribution
of the \dztokmn{} decay based on a sample of 13,000 events~\cite{focus}.
After subtracting known charm backgrounds, their $f_+(\qsq)$ distribution 
is an excellent match to a pole form with 
$m_\textrm{pole} = 1.91 \pm 0.04 \pm 0.05 \gevcsq$ or to 
a modified pole form with $\alpha = 0.32$.

BELLE (06) measured the absolute branching fractions and form factors
of the \dztokln{} and \dztopiln{} decays using both muonic and electronic
modes~\cite{belle}.
By fully reconstructing the recoil charm meson and the other mesons from
fragmentation, they ensured an excellent \qsq{} resolution and a low level of
backgrounds albeit with considerably reduced statistics. They obtained
2,700 \dztokln{} and 300 \dztopiln events from a 282 \fb data sample.
Their data set was fit to the simple pole form and the modified pole form.

BABAR (06) reported a preliminary measurement of the form factors of the
\dztoken{} decay~\cite{babar}. By applying the $D^*$ tag, they
obtained a sample of 100,000 events based on a 75 \fb data sample.
They also used the simple pole form and the modified pole form to fit their
data.

CLEO (06) reported preliminary measurements of the absolute branching
fractions and form factors of \dztoken{}, \dztopien{}, \dtokzen{}, and
\dtopizen{} decays based on a sample of 281 \pb~\cite{cleonadia}.
They did not require the recoil side $\overline{D}$ tagging, which enhanced 
the size of the selected events with a slight penalty in the \qsq{} resolution.
Both the simple and modified pole forms are used for fitting.

\begin{table}
\caption{Summary of recent experimental results on the form factor studies
of $D \to K \ell \nu_\ell$. Only the fit parameter $\alpha$ of
the modified pole form is shown here. CLEO-c obtained a low $\alpha$  while
BABAR and BELLE obtained high numbers compared with other experiments.
For other fit results and information
on $D \to \pi \ell \nu_\ell$ decays, check the reference for each
experiment.}
\label{alpha}
\begin{tabular}{l|l|l}
\hline
 Experiment & $\alpha$ & Channel\\
\hline
 CLEO III~\cite{cleo3} & $0.36 \pm 0.10_{\ -0.07}^{\ +0.08}$
                       & $D^0$ to $e$, $\mu$  \\
 FOCUS~\cite{focus} & $0.28 \pm 0.08 \pm 0.07$ & $D^0$ to $\mu$\\
 BELLE~\cite{belle} & $0.40 \pm 0.12 \pm 0.09$ & $D^0$ to $e$\\
                         & $0.66 \pm 0.11 \pm 0.09$ & $D^0$ to $\mu$\\
 BABAR~\cite{babar} & $0.43 \pm 0.03 \pm 0.04$ & $D^0$ to $e$\\
 CLEO-c~\cite{cleonadia} & $0.19 \pm 0.05 \pm 0.03$ & $D^0$ to $e$\\
                              & $0.20 \pm 0.08 \pm 0.04$ & $D^+$ to $e$\\
\hline
\end{tabular}
\end{table}

\section{FORM FACTORS OF VECTOR DECAYS}

The kinematics of $D \to \overline{K}^* \ell \nu_\ell$ decays is determined
by form factors and five kinematic variables:
\qsq{}, $m(K\pi)$, the lepton decay angle \thl, the vector decay angle
\thv, and the acoplanarity between two decay planes $\chi$.
For this type of decays, K\"orner and Schuler~\cite{ks} proposed
a set of helicity-basis form factors \Hpls, \Hmin, and \Hzer, which are,
in principle, computable by QCD theories. After integrating over $\chi$,
the decay intensity of the \dtokpiln{} process becomes proportional to
\begin{equation}
  \qsq \left[ \begin{array}{l}
  ((1+\costhl)\sinthv)^2 |\Hpls|^2 |BW|^2\\
  + ((1-\costhl)\sinthv)^2 |\Hmin|^2 |BW|^2\\
  + (2\sinthl\costhv)^2 |\Hzer|^2 |BW|^2\\
  + 8(\sinthl^2\costhv) \Hzer\hzer Re\{Ae^{-i\delta}BW\}\\
  + O(A^2) + m_\ell\ \textrm{terms}
\end{array} \right],\label{vector}
\end{equation}
where $BW$ and $Ae^{-i\delta}$ represent the $\overline{K}^*$ resonance and
the $s$-wave in the $K$-$\pi$ system, respectively. A new form factor \hzer{}
for the $s$-wave is introduced to describe the interference between the
$s$-wave and $\overline{K}^*$.
In the K\"orner and Schuler scheme, $H_\pm$ and $H_0$ are
linear combinations of two axial-vector and one vector form factors. 
In the past, it was assumed that the axial- and vector form factors take the
simple pole form such as $A_i(\qsq) = A_i(0)/(1-\qsq/M_A^2)$ and
$V(\qsq) = V(0)/(1-\qsq/M_V^2)$. With $M_V$ and $M_{A_i}$ fixed at 2.1 and
2.5 \gevcsq, experiments measured the ratio $R_V = V(0)/A_1(0)$ and
$R_2 = A_2(0)/A_1(0)$ from their data sets. Recently, following the idea of
Becirevic and Kaidalov~\cite{bec}, Fajfer and Kamenik~\cite{fk} proposed a
new parameterization for the vector semileptonic decays, where some
of the form factors are replaced by effective poles.

The world averages of the form factor ratios of the \dtokrln{} decays
are: $R_V = 1.66 \pm 0.060$ and $R_2 = 0.827 \pm 0.055$~\cite{pdg2004}.
The experimental results are based on the simple pole parameterization and
fairly consistent among themselves with small uncertainties.
However, there hasn't been any serious experimental check on the shape
of the form factors yet.

Based on their large data samples, FOCUS (05)~\cite{focuskpiln} and
CLEO (06)~\cite{cleokpiln} obtained details on the shape of the helicity form
factors. The functional form of Eq.~\ref{vector} shows that one can
disentangle the helicity form factors based on their different angular bin
populations. Monte Carlo simulations are used to calculate the projection
weights as a function of decay angles, The form factors \Hspls, \Hsmin, \Hszer,
and \Hint{} are obtained from projectively weighted histograms.

CLEO compared the shape of \Hspls, \Hsmin, \Hszer, and \Hint{} to a parametric 
model based on a FOCUS paper~\cite{focuspara} and found a good agreement except
the interference term. The presence of the $s$-wave interference is confirmed,
but there is a small deviation from the model in the \Hint{} shape.
More data would be needed to obtain detailed information on the shape of \hzer.
CLEO also compared their data to a model based on constant axial- and vector
form factors.  The data fits this model equally well,
implying little sensitivity to the simple pole masses with present statistics. 

\section{SUMMARY}

The field of the semileptonic charm decay is very active. Both inclusive
and exclusive branching fractions are updated recently. Several experiments are
involved in a healthy competition to measure charm semileptonic form factors
as accurately as possible. In light of recent experimental progress,
new ideas on fit parameterization ideas and lattice QCD calculations
are emerging.

The topics covered in this article is just part of the story. More updates
on exclusive branching fractions, rare decays, $D_S$ semileptonic decays
and form factor measurements will appear in near future.

\end{document}